\documentstyle[pre,aps,multicol,epsfig]{revtex}

\begin{document}
\draft

\title{First Experimental Evidence for Chaos-Assisted Tunneling \\
       in a Microwave Annular Billiard
      }
\author{C. Dembowski$^{1}$, H.-D. Gr\"af$^{1}$, A. Heine$^{1}$, 
        R. Hofferbert$^{1}$, H. Rehfeld$^{1}$, and A. Richter$^{1,2}$ 
       }
\address{$^{1}$
         Institut f\"ur Kernphysik, Technische Universit\"at Darmstadt,
         D--64289 Darmstadt, Germany \\
         $^{2}$
         Wissenschaftskolleg zu Berlin,
         D--14193 Berlin, Germany \\
        }

\date{\today}

\maketitle

\begin{abstract}
We report on first experimental signatures for chaos-assisted tunneling
in a two-dimensional annular billiard. Measurements
of microwave spectra from a superconducting cavity with high
frequency resolution are combined with
electromagnetic field distributions experimentally determined 
from a normal conducting twin
cavity with high spatial resolution to resolve eigenmodes
with properly identified
quantum numbers. Distributions of so-called quasi-doublet splittings
serve as basic observables for the tunneling between 
whispering gallery type modes localized to congruent, but distinct tori
which are coupled weakly to irregular eigenstates associated
with the chaotic region in phase space.

\end{abstract} 
\pacs{PACS number(s): 05.45.Mt, 41.20.-q, 84.40.-x} 

\begin{multicols}{2}

\narrowtext

For two decades a new kind of tunneling mechanism produces great
interest, since it demonstrates how the dynamical features
of a classical Hamiltonian system effect the behavior of its quantum
counterpart \cite{DavisHeller,Wilkinson,Gutzwiller}.
This so-called ``dynamical tunneling'' occurs whenever a discrete
symmetry of the system leads to distinct but symmetry
related parts of the underlying classical phase space. 
In contrast to the well-known barrier tunneling, 
dynamical tunneling only depends upon the probability
for such a quantum particle, although classically forbidden, 
to leave certain regions of phase space and travel
into others. This basically involves the coupling strength between
distinct phase space regions. 
In the special case of {\it two} symmetry related 
{\it regular} regions separated by a {\it chaotic} area in a mixed phase space, 
semiclassical quantization yields pairs of quantum states 
which are localized to the corresponding sets of congruent, 
but distinct tori. 
These so-called quasi-doublets show a very sensitive splitting behavior
which depends upon the coupling to irregular eigenstates
associated with the intermediate chaotic sea.
This non-direct, enhanced coupling of regular eigenmodes via chaotic ones
is what defines chaos-assisted tunneling 
\cite{BohTomUll,TomsoUllmo,Tomsoallein} in its original sense
\cite{Aberg}. 

The aim here is to demonstrate for the first time
that chaos-assisted tunneling can be
observed experimentally, even for a case where the size of the splitting is
several orders of magnitude below the typical mean level spacing
of the system. For this purpose we performed measurements
on superconducting as well as normal conducting microwave cavities
constituting a special family of Bohigas' annular billiard
\cite{BohiRing,DorFri3,HackeNoeck}. 
This system has been proven 
in very extensive and certainly also very accurate computer simulations,
especially in Refs. \cite{BohiRing,DorFri3},
to be a paradigm for 
chaos-assisted tunneling and provides  
access for experimental investigation.

The two-dimensional geometry of the annular billiard 
is defined by two circles of radius $r$ resp. $R$, the latter
being set to unity, and the center displacement or eccentricity $\delta$,
see left part of Fig.1. In the following, only the special one-parameter
family $r+\delta=0.75$ will be considered, since it provides all the
features which are relevant for chaos-assisted tunneling:
From the classical point of view the system shows a transition from
integrable ($\delta=0$) to mixed behavior ($\delta>0$), thus  
developing a growing chaoticity with increasing $\delta$.
Furthermore, the discrete reflection symmetry  leads
to congruent but classically distinct regions in phase space. 

\begin{figure}  
\centerline{\epsfxsize=8.6cm
\epsfbox{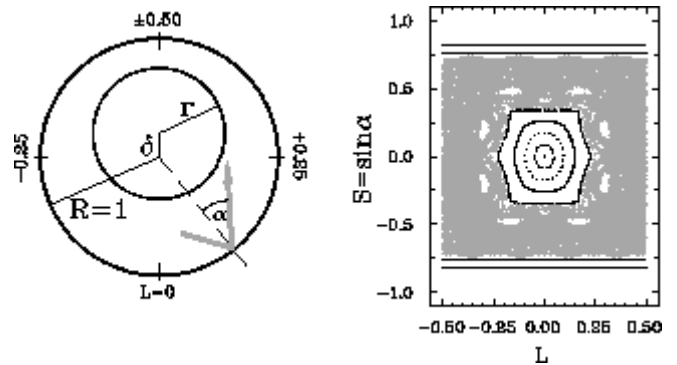}}
\vspace*{2.5mm}
\caption{
The annular billiard for $\delta=0.20$ and $r=0.55$ (left part)
together with the corresponding Poincar\'e surface of section (right part). 
Beside a large chaotic sea with stable islands in the center
the phase space clearly displays two symmetry related but dynamically
distinct regular coastal regions for $|S|>0.75$. 
Two examples of horizontal lines corresponding to 
whispering gallery trajectories 
(clockwise as well as counterclockwise) are shown.
}
\end{figure}

To demonstrate this, Fig.1 (right part) shows a typical Poincar\'e
surface of section for the configuration ($\delta=0.20/r=0.55$).
Here the area preserving Birkhoff coordinates $L$ (point of impact on the
the outer circle) and $S=\sin\alpha$ (angular momentum of the billiard
particle) have been used. Beside a large chaotic sea with chains of
stable islands in the center, both of which are influenced by a change of
the eccentricity
$\delta$, two symmetry related but distinct neutrally stable coastal regions
for $|S|>0.75$ can be observed. 
Per construction of the family $r+\delta=0.75$ these 
regular regions are invariant
under variations of $\delta$, since the corresponding trajectories
do not hit the inner circle. As a consequence $S$
is conserved, indicated by horizontal lines in the surface of section.
Those lines correspond to two so-called
whispering gallery trajectories \cite{BohiRing}.
From this, the only difference between the two distinct regular
regions is the sign of $S$, i.e. the sense of motion for the
propagating particle.

The fundamental question now accounts for the quantum counterpart
of the classically forbidden transport between the distinct coastal
regions: dynamical tunneling. Since the coupling between both
regions crucially depends upon the topology and the size of the chaotic sea,
the system is in particular adequate to study chaos-assisted tunneling.
But what is the basic observable for the tunneling strength in the
corresponding quantum system? To answer this it is very instructive
to start with the integrable case ($\delta=0$). Solving the Schr\"odinger
equation with Dirichlet boundary conditions leads to eigenvalues $k_{n,m}$
and eigenstates $\Psi_{n,m}$, with the
angular momentum quantum number $n$ and the radial quantum number $m$. Due to
EBK quantization \cite{Gutzwiller,BohiRing}
the property $S=n/k_{n,m}$ is the quantum angular momentum
which has to be compared with the classical $S=\sin\alpha$
in order to find the location of a certain quantum state in the 
phase space. While in the classical system the reflection
symmetry of the billiard leads to {\it two} distinct but related regular
regions with opposed sense of motion for the propagating
particle (i.e. the whispering gallery trajectories clockwise and
counterclockwise), the corresponding quantum eigenstates are organized 
in doublets for $\delta=0$ with {\it two} parities, even and odd, respectively.
However, continuously increasing the eccentricity $\delta$ 
systematically destroys
this doublet structure, yielding singulets for states with $S=n/k$
right within the chaotic sea ($S<0.75$) and quasi-doublets on the remaining
regular coast ($S>0.75$).
As in the case of the well-known double-well potential 
\cite{DavisHeller,Wilkinson,TomsoUllmo}
the very small splitting of those quasi-doublets is directly determined
by the classically forbidden tunneling, thus presenting a very
effective observable for the hardly accessible tunneling strength.
Since the location of a certain quasi-doublet on the regular coast
(defined by $S=n/k$) as well as the transport features of the chaotic
sea (defined by the eccentricity $\delta$) have a direct impact on the
splitting, its systematic investigation allows the experimental study
of chaos-assisted tunneling in the annular billiard.

As in earlier studies (for an overview, see \cite{Playing}), we 
simulated the quantum billiard by means of a two-dimensional 
electromagnetic microwave resonator of the same shape
(see l.h.s. of Fig.1). The measurements were divided into two parts:
Taking in total three different configurations of the family $r+\delta=0.75$
(i.e. $\delta=0.10$, 0.15 and 0.20) we performed on one hand
experiments with a superconducting Niobium resonator 
(scaled to $R=\frac{1}{8}$ m) at 4.2 K in order
to measure quasi-doublet splittings within the frequency range up to $f=20$ GHz.
The very high quality factor of up to $Q\approx 10^6$ allows
a resolution of $\Gamma/f\approx 10^{-6}$, where $\Gamma$ is equal to the
full width at half maximum of a resonance.
For demonstration, Fig.2 shows transmission spectra of the 
superconducting resonator in the vicinity of 9 GHz. Beside several
singulets exactly one quasi-doublet can be observed which 
shows a small but systematic displacement with 
the eccentricity $\delta$. 
In all cases the very small splitting of the quasi-doublet
is clearly detectable. This is only due to the
high frequency resolution of the superconducting resonator.
In this context, however, it is important to note, that the position
of the exciting antennas has to be choosen very carefully in order to minimize
the perturbation on the whispering gallery type 
modes in the coastal region (see Fig.1)
and thus not to influence the size of their physical
quasi-doublet splitting. Using antennas right within the whispering
gallery region of the billiard (see sketch on top of Fig.2)
always produce ``false'' splittings even
for the concentric system ($\delta=0$) with twofold degenerate states.
We therefore used antennas in the shadow region of the inner circle
also preserving the symmetry of the whole geometry, cf. Fig.2.

\begin{figure}  
\centerline{\epsfxsize=8.6cm
\epsfbox{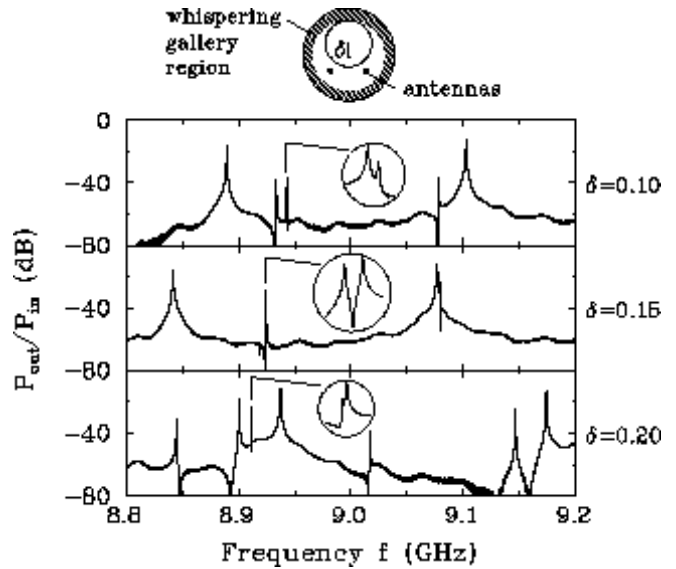}}
\vspace*{2.5mm}
\caption{
Transmission spectra around 9 GHz with varying eccentricity $\delta$.
Among several singulets exactly one quasi-doublet,
slightly moving with $\delta$, can be observed. In the zooming circles
the abszissa is stretched by a factor of 50 in order to visualize
the quasi-doublet splitting. Spectra  at eccentricites $\delta < 0.10$ could 
not be realized with the present set up of antenna locations.
}
\end{figure}

For a proper identification of the quantum numbers $(n|m)$ of the modes
associated with the quasi-doublets on the other hand
we used a normal conducting Copper twin of the Niobium billiard cavity.
There we measured the corresponding wavefunction resp. electromagnetic
field distributions
from which the relevant quantum numbers $n$ and $m$ could be deduced,
even if they are far from being ``good quantum numbers'' 
in the given eccentric systems which are non-integrable.
This second part of the experiment was based on a field perturbation
method originally introduced in accelerator physics and used
successfully in billiard research before \cite{Sridhar,Gokirmak}. 
According to
Slater's theorem \cite{Slater} a small metallic body inside the cavity
locally interacts with the electromagnetic field
in such a way that a frequency shift of the excited mode results from
the compensation of the non-equilibrium between the totally stored electric
and magnetic field energy. This frequency shift
\begin{equation}
\partial f=f_0-f=f_0\bigg(a\vec{E_0}^2-b\vec{H_0}^2\bigg)
\end{equation}
with respect to the unperturbed mode (index 0) directly depends
upon the superposition of the squared electric and magnetic fields,
$\vec{E_0}$ and $\vec{H_0}$, respectively. 
Since the quantum wavefunction is related to the electric
field only, the magnetic component has to be removed by a proper
choice of the geometry constants $a$ and $b$ in Eq.(1) by choosing needle-like
bodies (1.84 mm in length and 1.00 mm in diameter). 
Moving the body across the whole two-dimensional
surface of the billiard 
with a spatial resolution of about one tenth of a wavelength
by means of a guiding magnet
and detecting the frequency shift $\partial f$
at each position, finally provides the complete field distribution.
Examples of those for the configuration $\delta=0.20$ in the vicinity
of 9 GHz are plotted in the upper part of Fig.3. 

Of the three distributions
only the middle one with the quantum numbers $n=18$ (36 field maxima in the
polar direction) and $m=1$ (one field maximum in the radial direction) is
characteristic for the modes which are localized in the whispering
gallery region (see Fig.2). Contrary to this the distributions on the 
l.h.s. and on the r.h.s. show a totally different pattern. The parity
of the distributions is determined in the following way: If there is
maximum field strength on the line which defines the reflection symmetry
of the billiard, positive parity can be assigned to the mode, likewise
negative parity for zero field strength on this line. 

Underneath the squared electric field strength distributions in 
Fig.3, the corresponding transmission spectrum taken at 300 K with the
normal conducting Copper cavity is shown. The three broad resonances
associated with the field distributions are much better resolved in the 
measurement at 4.2 K of the superconducting Niobium twin cavity. The
small displacement in frequency of the resonances in the two measurements
is due to mechanical imperfections of each individual cavity and 
positioning errors of the respective inner circles within the resonators.
Mechanical uncertainties of order $\pm 100$ $\mu$m relative to the radius
of the outer circle $R=$125 mm are sufficient to account for the
observed displacements.

The spectrum at 4.2 K in Fig.3 shows that the resonance magnified
in the insert is in fact one of the expected quasi-doublets
characteristic for chaos-assisted tunneling. Naturally, this quasi-doublet
is not resolved in the spectrum taken at room temperature, and the field
distribution in the upper part of Fig.3 proves that of the two modes
corresponding to the doublet in the particular case considered the one
with negative parity is excited stronger than the other.

Calculating the quantum angular momentum $S=n/k$ 
(with $k=2\pi R/\lambda=2\pi R f/c_0$, where f denotes the centroid 
frequency of the
quasi-doublet and $c_0$ the speed of light) finally yields
the position of the whispering gallery type mode on the corresponding
classical surface of section, see r.h.s. of Fig.1.
In the case of mode $(18|1)$ one obtains $S\approx 0.77$ characteristic
for a mode in the so-called ``beach region'' \cite{DorFri3} defined by the
borderline $S=0.75$ between the chaotic sea and the regular coast,
respectively.

\begin{figure}  
\centerline{\epsfxsize=8.6cm
\epsfbox{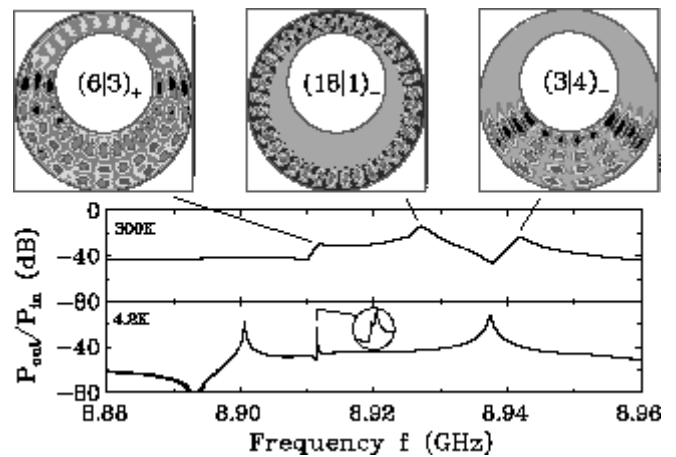}}
\vspace*{2.5mm}
\caption{
Matching of field strength plots taken at 300 K with high spatial resolution
and microwave spectrum taken at 4.2 K with high frequency resolution.
The combined normal conducting/superconducting setups allow
to measure highly resolved quasi-doublets including quantum numbers
$(n|m)$ and parity. See text.
}
\end{figure}

This comparison demonstrates that the 
measurements combine the high spatial resolution of about $\lambda/10$
for the normal conducting billiard with the high 
frequency resolution of about $1/Q$ for the superconducting one, 
thus allowing a very
effective classification of regular quasi-doublets as well as 
chaotic singulets in the range up to approximately 14 GHz, where the splittings
become smaller than the resonance widths of the superconducting
resonator. The difference in frequency between the peaks of each quasi-doublet
were estimated through a non-linear fitting to ``skew Lorentzians'' (see 
Eq.~(4) in \cite{brentano}). 

In what follows, we only consider the family with quantum numbers
$(n|1)$, since it consists of some 30 resolved and 
undoubtedly identified quasi-doublets within the
measured frequency range. 
To uncover effects due to chaos-assisted
tunneling, the splitting of a certain quasi-doublet has to be analyzed
as a function of its corresponding position in the classical phase space.
As mentioned above, this position might be expressed in terms of 
$S=n/k$ as the quantized analog of the classical angular momentum 
$S=\sin\alpha$, directly representing the very location in the underlying
surface of section. The resulting curve, again for the configuration 
$\delta=0.20$, is shown in the lower part of Fig.4.
\begin{figure}  
\centerline{\epsfxsize=8.6cm
\epsfbox{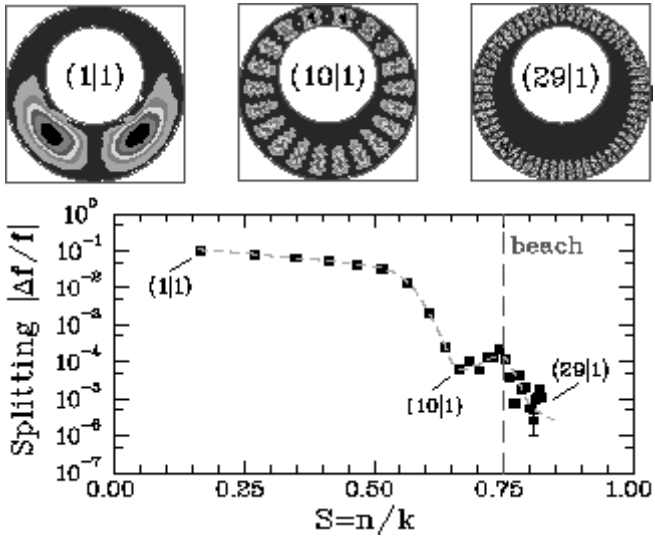}}
\vspace*{2.5mm}
\caption{
Distribution of normalized splittings with respect to the position
of a certain quasi-doublet in classical phase space. The error on the splitting
is generally smaller than the size of the squares except for the smallest
splitting observed where an error bar is placed on the point. Beside
a smooth transition from states with a large splitting right within
the chaotic sea ($S<0.75$) towards states with a very small splitting
on the regular coast ($S>0.75$) a local maximum occurs in the direct 
vicinity of the beach, representing a very impressive signature
for chaos-assisted tunneling.
}
\end{figure}

Here, the distribution of normalized splittings
$|\Delta f/f|$ shows a very smooth transition from chaotic states
defined by large splittings right within the chaotic sea ($S<0.75$)
to regular quasi-doublets with very small splittings in the classical
coastal region ($S>0.75$). Also the measured field distributions show an
increasing regularity with growing $S$, as can be seen
from the examples in the upper part of Fig.4. 
Especially mode $(29|1)$ is hardly distinguishable from 
the corresponding concentric mode (with splitting zero) although the
system is strongly eccentric.

Besides this global behavior a first strong signature of 
chaos-assisted tunneling can be observed in the particular shape 
of the splitting curve:
In the direct vicinity of the beach at $S=0.75$ the quasi-doublets
show a locally enhanced splitting amplitude, thus indicating a very effective
coupling between the regular coast and the chaotic sea. As described above, 
this corresponds to a locally enhanced tunneling strength in the beach
region as theoretically predicted in \cite{DorFri3}.
\begin{figure} 
\centerline{\epsfxsize=8.6cm
\epsfbox{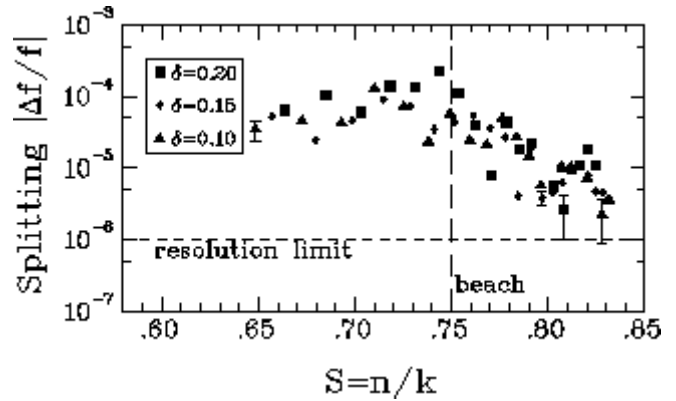}}
\vspace*{2.5mm}
 \caption{
Distribution of normalized splittings in the zoomed beach region 
for different eccentricities $\delta$.
The local maximum is rebuilt 
by all measured quasi-doublets forming a systematically rising and a 
randomly falling part below and above $S=0.75$, respectively.
}
\end{figure}
To evaluate the influence of the chaoticity of the system,
Fig.5 shows a direct comparison of the splittings for all
eccentricities $\delta$ in the vicinity of the beach at
$S=0.75$. Note that the splittings for different $\delta$ 
not only enhance the visibility of the maximum, they furthermore
reveal an additional feature of chaos-assisted tunneling:
While splittings on the rising left part of the maximum, i.e.
within the chaotic sea ($S<0.75$), are distributed quite systematically
--- e.g. the data points of $\delta=0.20$ always correspond to the
largest splittings --- shows the falling right part large fluctuations
for a given eccentricity $\delta$. This effect is also theoretically 
predicted \cite{TomsoUllmo,BohiRing,DorFri3} and accounts
for the high rate of anti-crossings with chaotic modes for high angular
momenta $S$. Thus on the r.h.s. of $S=0.75$
the tunneling strength shows a very random dependence
on the eccentricity $\delta$ leading to strong fluctuations in the 
distribution of splittings. 
Finally, for even larger values of $S$ the splitting amplitudes
are of the order of the inverse quality factor, 
$\Delta f/f\approx 1/Q\approx 10^{-6}$ defining the resolution limit of
the present setup.

In summary, we have presented first experimental signatures for 
chaos-assisted tunneling in a billiard. As the basic observable
we have investigated the splittings of quasi-doublets with respect
to their position in the classical phase space and their dependence
on the eccentricity $\delta$. 
A local maximum in the vicinity
of the beach region with a systematically rising and randomly
falling part has been found which directly reflects the enhanced
tunneling strength at this critical location between the regular coast
and the chaotic sea. 
In this context, a combined experimental setup using normal as well as
superconducting billiards has offered a very effective tool for measuring
highly resolved quasi-doublets with properly identified quantum numbers.

We are particularly grateful to O. Bohigas for encouraging us to study
this novel mechanism of tunneling and him,
S. Tomsovic and D. Ullmo for their kind invitations to Orsay 
and many fruitful discussions. 
One of us (A.R.) has also profited very much from M.C. Gutzwiller's
insight. We thank E. Doron and S. Frischat
especially for guiding us to ``look on the beach''.
This work has been supported by the DFG under contract number 
Ri 242/16-1 and through the SFB 185 ``Nichtlineare Dynamik''.

\end{multicols}

\end{document}